# An Interactive Learning Tutorial on Quantum Key Distribution

Seth DeVore and Chandralekha Singh

*Department of Physics and Astronomy, West Virginia University, Morgantown, WV 26506*
*Department of Physics and Astronomy, University of Pittsburgh, Pittsburgh, PA 15260*

**Abstract:** We describe the development and in-class evaluation of a Quantum Interactive Learning Tutorial (QuILT) on quantum key distribution, a context which involves an exciting application of quantum mechanics. The protocol used in the QuILT described here uses single photons with non-orthogonal polarization states to generate a random shared key over a public channel for encrypting and decrypting information. The QuILT strives to help upper-level undergraduate students learn quantum mechanics using a simple two state system. It actively engages students in the learning process and helps them build links between the formalism and the conceptual aspects of quantum physics without compromising the technical content. The in-class evaluation suggests that the validated QuILT is helpful in improving students' understanding of relevant concepts.

## 1. INTRODUCTION

Quantum mechanics is a particularly challenging subject for undergraduate students and several investigators have focused on diverse topics, e.g., on visualizing potential energy diagrams, hands-on activities integrated with technology for teaching quantum mechanics, spontaneous models of conductivity, one -dimensional scattering, single photon experiments in undergraduate labs, representations for a spins-first approach, structural features of quantum notation, quantum curriculum based upon simulations, energy measurement and time-dependence in quantum mechanics, epistemological framing in quantum mechanics and student understanding of wavefunction in asymmetric wells [1-16]. We have been using research on student difficulties in learning quantum mechanics as a guide to develop learning tools, e.g., a set of Quantum Interactive Learning Tutorials (QuILTs) on concepts covered in an upper-level quantum mechanics course, e.g., time-dependence of wave function, Stern-Gerlach experiement, quantum measurement, addition of angular momentum, Dirac notation, Larmor precession of spin, observables and their corresponding quantum mechanical operators, probability distribution for measuring different observables, expectation values of observables, double slit experiment, Mach-Zehnder Interferometer with single photons, degenerate perturbation theory including use of perturbation theory for finding the energy level splitting due to the fine structure of hydrogen atom and Zeeman effect, and identical particles that strive to help students develop a good grasp of quantum mechanics [17-48]. The QuILTs use an inquiry-based approach to learning in which students are asked a series of guiding questions and they strive to bridge the gap between quantitative and qualitative aspects of learning quantum mechanics. The instructors can use the QuILTs either as in-class tutorials on which students can work in small groups or as homework supplements. Here, we discuss the development and evaluation of a research-validated QuILT related to quantum cryptography on quantum key distribution (QKD) [49-58] using two non-orthogonal polarization states of photons. We note that we have developed and validated another QuILT on QKD involving two entangled spin-1/2 particles which is also comparable in difficulty for undergraduate students based upon one-on-one interviews we have conducted since it involves entanglement between particles. We are still in the process of collecting in-class data so the findings from that QKD QuILT based upon entangled spin-1/2 particles will be presented elsewhere.

QKD is an interesting application of quantum mechanics useful for generating a shared secure random key for encrypting and decrypting information over a public channel [49-58]. QKD is already in use by the banking industry [56]. This real world application makes QKD a particularly useful vehicle for helping students learn about the fundamentals of quantum mechanics in an undergraduate course. A unique feature of secure QKD protocols is that two parties, e.g., a sender who is traditionally referred to as Alice and a receiver traditionally referred to as Bob, who generate a random shared key over a public channel, can detect the presence of an eavesdropper (traditionally referred to as Eve) who may be intercepting their communication during the shared key generation process to gain access to the key. Alice and Bob are connected via a quantum communication channel which allows quantum states to be transmitted during the random shared key generation process. When photon polarization states are used for shared key generation, the quantum communication channel used is generally an optical fiber or free space. In addition, during their shared key generation process, Alice and Bob also communicate certain information with each other via a classical channel such as the internet.

The protocol discussed here that students learn via the QKD QuILT involves generating a shared key over a public channel using single photons with non-orthogonal polarization states. The warm-up for the QuILT helps students learn background topics including the fact that in a two state system, a quantum state can be in a superposition of two linearly independent states but measurement of an observable will collapse the state and one obtains a definite value for the observable measured. In the first part of the QuILT after the warm-up, students learn about an insecure QKD protocol in which two orthogonal polarization states of single photons are used. Then they learn about the secure QKD protocol which uses two non-orthogonal polarization states of single photons via a guided inquiry-based approach to learning in which students are actively engaged in the learning process.

In the next section, we review the QKD protocol students learn via the QKD QuILT. We then describe the methodology, structure of the QKD QuILT, qualitative details of student difficulties, quantitative results from the pretest, posttest, and a delayed posttest and then summarize the results. The pretest was administered after traditional instruction about the basic quantum mechanics required for QKD but before students worked on the QuILT. The posttest was administered after the QuILT and the delayed posttest was administered roughly two months after the QuILT.

## 2. THE QKD PROTOCOL STUDENTS LEARN

A classical key distribution protocol cannot guarantee that the key shared over a public channel is secure. Such a classical protocol depends on the computational difficulty of mathematical functions. On the other hand, in order to securely generate a random shared key for encrypting and decrypting information over a public channel, QKD uses properties of quantum states and involves encoding information in quantum states. When measurement is performed in order to gain information, the state of the system collapses and one "bit" of information is obtained. In particular, the QKD protocols involve preparing and sending special quantum states and measuring an observable which yields one of two outcomes (either bit 0 or 1) in the prepared states with a certain probability. Though eavesdropping is possible for the secure quantum key generation protocol that students learn, comparing a small subset of the bits in the key at the end of the shared key generation process will reveal if an eavesdropper was present and if so, to what extent the key was compromised.

The ability to detect an eavesdropper in secure QKD protocols is due to the fact that physical observables are in general not well-defined in a given quantum state but measurement collapses the state and gives the observable a definite value. In particular, secure key generation exploits the fact that in general, measurement disturbs the state of a quantum mechanical system and that a random unknown quantum state cannot be cloned [58]. The indeterminacy is unique only to quantum mechanics and can be used to determine if someone eavesdropped during the QKD process and if so, how much information was gained by the eavesdropper, Eve. In particular, Eve must measure an observable in the state of the system sent by Alice to Bob during the shared key generation process over the public channel. But since quantum measurement in general disturbs the state of the system sent by Alice without guarantee that the state will be identified by the person performing the measurement, Eve will not always know the state sent by Alice. Even if Eve intercepts what Alice transmitted and uses clever strategies for sending replacement quantum states, since she must guess the state to transmit to Bob at least in some cases, it will cause a discrepancy in the shared key that Alice and Bob generate in her presence. Also, if Bob is alerted by Alice that she has transmitted a state in the key generation process, and the eavesdropper does not send a replacement state to Bob to replace what Alice had transmitted, the discrepancy present in the shared key will be such that Alice and Bob will detect the presence of an eavesdropper. After the entire shared key is generated over the public channel, Alice and Bob can compare certain bits, e.g., every $p^{th}$ bit, to ensure that they agree on the shared key generated (they discard the bits they compare so that it is not part of the shared key). If they find discrepancy in the bits they compare (beyond a threshold discrepancy due to decoherence in actual situations), they will abort the shared key generated because someone may be eavesdropping.

In the QKD protocol Alice randomly sends a series of single photons with either a +45° or 0° polarization. Bob examines each photon by passing it through a polarizer with a polarization axis of either -45° or 90° with a photodetector set up behind the polarizer which clicks if the photon passes through his polarizer. In the two possible cases in which the photon's polarization is not orthogonal to the polarization axis of Bob's polarizer, there is a 50% chance for the photon to pass through and a 50% chance for the photon to be absorbed. In the other two possible cases, in which the photon's polarization is orthogonal to the polarization axis of Bob's polarizer, there is a 100% chance for the photon to be absorbed. Therefore, on average, 25% of the time the photon will pass and be detected. Since the photon must not be polarized orthogonal to the polarization axis of Bob's polarizer for it to pass through, Bob can determine the polarization of the photon when it passes through the polarizer and is detected by the photodetector behind the polarizer. If the photon is not detected by the photodetector, it could be polarized along either of the two possible directions (though it is more likely that it is polarized orthogonal to the polarization axis of Bob's polarizer) and Bob cannot be certain of the polarization of the photon. Alice sends Bob photons at regular intervals (e.g., once per millisecond). The distance is known, and so Bob knows when to expect each photon. If the photon is not detected, neither Alice nor Bob record the next bit in the key and Alice sends another photon and contacts Bob again. If a photon is detected, Bob confirms that he has detected the photon and both he and Alice record the next bit in the key (e.g., they agree ahead of time that if Bob detects +45° polarized photon that Alice had sent, they will record +1 and if he detects 0° polarized photon she sent, they will record a 0 for the shared key). If we introduce an eavesdropper (Eve) who is using the same protocol as Bob to intercept photons sent by Alice and replaces them to the best of her ability, she (like Bob) will not be certain of the polarization of photon that she intercepted 75% of the time (when her detector does not register the photon). Even if Eve makes the best guess for all cases in which she is not certain and assumes that the photon is polarized orthogonal to the polarization axis of her polarizer which will be correct in 2/3 of these cases, she will send replacement photons with the incorrect polarization of photon 25% of the time resulting in a discrepancy between Alice and Bob's key (which can be detected by Alice and Bob if they compare a subset of the total key over a public channel after the entire key is generated).

Before learning about the QKD protocol for secure key distribution involving two non-orthogonal polarization states of single photons, students learn about an *insecure* key distribution protocol using two orthogonal polarization states of single photons. In the insecure protocol, Eve can find out the polarization states of each photon that Alice sends to Bob while generating the shared key over a public channel with 100% certainty without her presence being detected by Alice and Bob. In particular, students learn that an eavesdropper in the insecure QKD protocol can generate a replacement photon with the same polarization as the one sent by Alice to Bob during the shared key generation process and transmit it to Bob without him realizing that the photon he intercepted did not come from Alice but came from Eve. The QuILT helps students learn that the ability for an eavesdropper to gain access to the key without her presence being detected makes this protocol insecure for generating a shared key over a public channel.

The QKD QuILT guides students through an ideal situation in the absence of decoherence. However, students learn that error can be introduced in the shared key generated by Alice and Bob due to decoherence as well (e.g., interaction of the photon with the surroundings which can change its polarization state or lead to scattering or absorption of the photon). They learn that the effect of decoherence can be neglected over small distances, e.g., QKD schemes, similar to the one they learn have been tested successfully and the highest bit rate system currently demonstrated exchanges secure keys at 1 Mbit/s (over 20 km of optical fiber) and 10 kbit/s (over 100 km of fiber) with applications in the banking industry [56].

## 3. METHODOLOGY

**Participants and Course Details:** The participants in this study were undergraduate students enrolled in the first semester of a two-semester upper-level quantum mechanics course at a large research university in the US serving more than 30,000 students. The first semester of this two-semester sequence in which this study was conducted is mandatory for all physics majors. A majority of these students are juniors and seniors (typically greater than 75% are seniors). There are typically 20% women in this course. The course instructor selected the textbook by David Griffiths on quantum mechanics as the textbook from which most of the quantitative homework problems assigned each week were selected. The students examined in this analysis included all students enrolled in this course during the 2009, 2010, and 2013 years. The instructor also used QuILTs and the corresponding pretests and posttests on various topics throughout the course. The instructor would typically teach a topic via traditional lecture, then give students a pretest in class to gauge student understanding of those concepts after traditional lecture, then students would work in small groups in class (or sometimes extended over two class periods) on the corresponding QuILT. If the students could not finish the QuILT in class and another class period was not allocated for it, they were asked to finish the rest of that QuILT as part of the homework. The students were given the posttest after they had completed the corresponding QuILT. The pretest on a topic (which was after traditional instruction in relevant concepts) and posttest (after the associated QuILT on that topic) both counted as quizzes but the pretest was graded for completeness while the posttest was graded for correctness. The final version of the questions used in the pretest are available in the Appendix (posttest questions were similar but the angles were different).

In the first semester of quantum mechanics, the topic of QKD was covered after students had already learned about the basic postulates of quantum mechanics, time-dependence of states, quantum measurement etc. both in the context of infinite dimensional Hilbert space and for finite dimensional vector space, e.g., for a two-state system. Students also learned about addition of angular momentum after learning about spin-1/2 and other finite dimensional quantum systems. Students were also introduced to photon polarization states as an example of a two-dimensional Hilbert space and then they learned about the QKD via lecture-based instruction. We note that even though the instructor used the QuILT on Mach-Zehnder interferometer (MZI) with single photons and polarizers also involving photon polarization states, the MZI with single photons was introduced after the students had already engaged with the QKD QuILT since MZI with polarizers involves product space of path and polarization states. Thus, QKD QuILT with two non-orthogonal photon polarization states was the first QuILT involving photon polarization states that the students engaged with.

**Methodology for QKD QuILT Development and Validation**: The development and validation of the QKD QuILT (which includes a pretest and posttest as well as a warmup covering the required basics of photon polarization states and then the main tutorial) began with an analysis of student difficulties with related concepts after lecture-based instruction and how to help students develop a better understanding of these concepts. It builds on students' prior knowledge found via investigation of difficulties and uses a guided inquiry-based approach in which various concepts build on each other gradually. The development and validation of the QuILT went through a cyclic iterative process which included the following stages:

(1) Development of the preliminary version based on a theoretical cognitive task analysis [61] of the underlying knowledge and research on student difficulties with relevant concepts.
(2) Implementation and evaluation of the QuILT by administering it individually to students and getting feedback from faculty members who are experts in these topics.
(3) Determining its impact on student learning and assessing what difficulties were not adequately addressed by the QuILT.
(4) Refinements and modifications based on the feedback from the implementation and evaluation.

Individual interviews with 15 students were carried out using a think-aloud protocol [59] to better understand the rationale for their responses throughout the development of various versions of the QuILT and the development of the corresponding pre-test and post-test given to students before and after they engaged in learning via the QuILT. The students interviewed were from the undergraduate quantum mechanics course or graduate students (who had taken first-year graduate level core quantum mechanics course) who volunteered to aid in the development after an announcement made in the undergraduate quantum mechanics course and to graduate students. The 12 undergraduate students had lecture-based instruction on photon polarization states as well as on QKD. The 3 graduate students had learned about the polarization states of photon but not necessarily QKD and were interviewed in the later part of the QKD QuILT development when it only needed small refinement to understand the extent to which graduate students who had basic knowledge of the polarization states of photon could benefit from using the QuILT as a self-study tool. The interviewed undergraduate students typically had varying performance in the course up to the point the interview was conducted and even the graduate students who volunteered to interview had varying performance in their core quantum mechanics course. They were paid $25 for their time which was between 1 to 2 hours.

During the semi-structured interviews, students were asked to verbalize their thought processes while they answered the questions about the QKD basics (after traditional lecture-based instruction in relevant concepts) either as separate questions before the preliminary version of the QuILT was developed or as a part of the QuILT, which included various protocols for generating insecure or secure shared key over a public channel. If students became quiet for some time they were asked to keep talking, but were otherwise not interrupted. After the students answered the questions to their satisfaction, we asked them for clarification of the issues they had not made clear earlier. In the interviews that were conducted before the development of the QuILT, students were asked general questions relevant for the basics of the QKD QuILT. In later interviews conducted with different versions of the QuILT, students were asked to work on different parts of the QuILT including the basics while thinking aloud. In addition, students were also asked open-ended questions about basic issues relevant for QKD as they came up. Since the interviews were typically one hour long and the earlier interviews were more exploratory and consumed significant time investigating student understanding of the QKD basics, interview time was optimized by asking the basics of QKD questions only to the first five students and later interviews focused more extensively on these issues specifically in the context of different aspects of the QKD protocols (e.g., the non-secure protocol using orthogonal polarization states of light and secure protocol using non-orthogonal polarization states of light).

The QKD QuILT asks students to predict what should happen in a particular situation and after their prediction phase is complete, they are provided a figure and table from which they can infer what they should have predicted. Then, they are asked to reconcile the differences between their prediction and what they infer from the information provided. After each individual interview with a student with a particular version of the QKD QuILT (along with the administration of the pre-test and post-test to that student), modifications were made based upon the feedback obtained from students. For example, if students got stuck at a particular point and could not make progress from one question to the next with the hints already provided, suitable modifications were made to the QuILT. We also iterated all components of the QKD QuILT with three faculty members and made modifications based upon their feedback.

**Methodology for QKD QuILT in Class Implementation:** When we found that the QKD QuILT was working well in individual administration to students in one-on-one interviews and the post-test performance was significantly improved compared to the pre-test performance, the QuILT was administered in class to the undergraduate students in the quantum mechanics course for three years when the same instructor taught the course. Students worked on the QKD QuILT in class in small groups. All students in these quantum mechanics classes are included in our data. Similar to all of the other QuILTs the instructor implemented, before students worked on the QKD QuILT, the instructor lectured about the polarization states of photons and presented a basic outline of the QKD in the QuILT. Then, the students were administered the QKD pretest in class, which consists of the questions presented in the Appendix and individually discussed in the results section, and then they were given the full credit for the quiz for trying their best on the pretest (similar to all other QuILTs). The QKD QuILT was then administered to the students in class and they were allowed to spend the remainder of the class working on the QuILT in small groups of 2-3. Since most student groups did not complete the QuILT during the class, they were allowed to take the QKD QuILT with them to complete it as homework. On average, students noted spending around 1.5 hours on the entire QKD QuILT including the in class time. The posttest was administered after all students had the opportunity to complete the QuILT and submitted it as homework. The posttest was counted as a quiz and graded for correctness. The pretest and posttest are identical in structure with only the polarization angles changed between the pretest and posttest. The pretest and posttest each begin by briefly outlining a situation similar to the QKD protocol to ensure that the students are familiar with it. The pretest (and posttest) then asks a series of questions regarding a QKD protocol setup with Alice's polarization axes randomly switched between 70° or 0° (60° or 0° for the posttest) and the polarizer axes for Bob randomly switched between -20° or 90° (-30° or 90° for posttest). All pretests and posttests were graded by graduate researchers working on the QuILT and the resulting averages on each pre and posttest problem were recorded. The *t*-tests were performed on the posttest data to look for statistical differences between them.

The delayed posttest was given to students roughly two months after the QuILT in one of the three classes (the other two classes were not given delayed posttest). It was administered as a part of the final exam for the class and was comprised of two questions that were similar in structure to two questions present on the pretest and posttest with the only difference being the angles used. Since the instructor did not have time to administer more than two questions as the delayed posttest, only two questions were selected. For the delayed posttest Alice's polarization axes randomly switched between 50° or 0° and the polarizer axes for Bob randomly switched between -40° or 90°.

## 4. THE STRUCTURE OF THE QKD QUILT

The entire QKD QuILT is available here [60]. The QuILT strives to provide enough scaffolding to allow students to build a good knowledge structure while remaining engaged in the learning process. The benefit of the guided aspect of the QuILT is illustrated by one interviewed student noting that even if he did not know it earlier "[the answers to] most of these [questions] I can figure out if I just think about it." While this is just one students statement it is illustrative of other similar comments made by several of the students who explicitly noted that the guided approach helped them learn.

The QuILT warm up first helps students learn the basics related to bits, qubits, polarization states of a photon, and effect of measurement on a two-state system. Many of these basics are asked in a series of multiple-choice questions in a guided approach to learning. The need for reinforcing these basics in a guided approach was evident from answers to written free-response questions and during individual interviews with students. For example, after going over the basics, one interviewed student noted "If I hadn't known the stuff in the basics, I would be really confused with the QKD part" which was a relatively common sentiment of interviewed students regarding the usefulness of the basics. The following questions (not necessarily consecutive) in the warm up section (basics section) help students learn about a qubit:

**Q**: The quantum-mechanical analogue of a classical bit is called a "qubit". A qubit can be described as $|q\rangle = \alpha|0\rangle + \beta|1\rangle$ where $|0\rangle$ and $|1\rangle$ are two orthonormal states of a quantum system and $\alpha$ and $\beta$ are two complex numbers such that $|\alpha|^2 + |\beta|^2 = 1$. Let $\hat{Q}$ be an operator corresponding to a physical observable such that $\hat{Q}|0\rangle = 0|0\rangle$ and $\hat{Q}|1\rangle = 1|1\rangle$. Choose all of the following statements that are correct:

(I) A bit cannot be in both $|0\rangle$ and $|1\rangle$ states. However, a qubit can be in a superposition of $|0\rangle$ and $|1\rangle$ states.
(II) A measurement of observable $Q$ in a state $|q\rangle$ can only produce one of two possible values.
(III) When you perform a measurement of $Q$ in a qubit, you obtain one bit of information.

**Q**: Choose all of the following that can form a qubit (i.e., that can be used as basis states for a qubit):

(I) the eigenstates of $\hat{S}_z$ for a spin one-half particle
(II) the two orthogonal polarization states of a photon
(III) the ground state and the first excited state of a one-dimensional infinite square well.

We find that discussing these kinds of questions with other students while working on the QuILT in small groups helps students grasp the basic concepts that are necessary to understand how quantum key distribution works and why a particular protocol is secure or insecure. The following question is also asked in the warm up after a series of questions about specific polarization states of a photon:

**Q**: If a single photon with a normalized polarization state $|S\rangle = \alpha|H\rangle + \beta|V\rangle$ is incident on a horizontal polarizer, which one of the following is true?
(a) The photon passes through the polarizer with a probability $|\alpha|$.
(b) The photon passes through the polarizer with a probability $|\beta|$.
(c) The photon passes through the polarizer with a probability $|\alpha|^2$.
(d) The photon passes through the polarizer with a probability $|\beta|^2$.

Once students work through the basics, they are first led through the *insecure* protocol using orthogonal polarization states of photons. They learn that the protocol is not secure due to the use of orthogonal polarization states and eavesdropping will go undetected. In this protocol, students are led through a series of questions without the eavesdropper. Then, they are asked about Eve's interference. Students are guided to realize that Eve is capable of intercepting and replacing photons without her presence being detected. This insecure protocol gets students used to the process by which a shared key can be generated but illustrates a flawed method that is prone to eavesdropping going undetected.

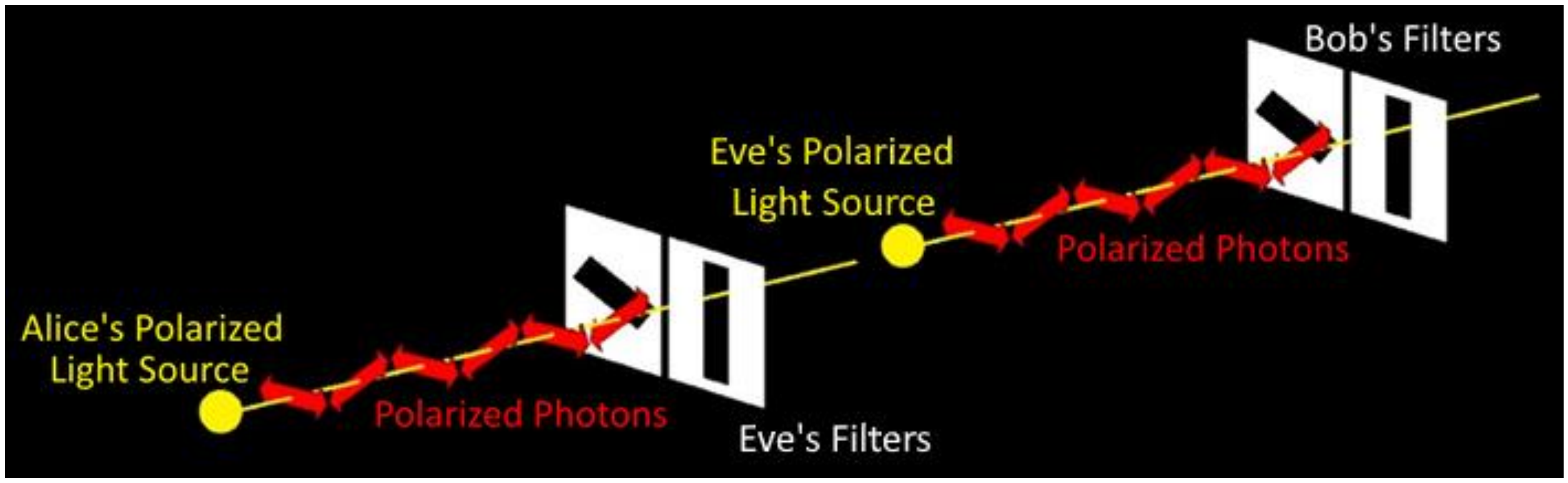

**Figure 1.** Eve's interference absorbs the original photon produced by Alice and requires her to generate a replacement photon to send to Bob to avoid detection.

Next, students are guided through the QKD protocol which generates a secure key in which Alice, Bob and Eve each use two non-orthogonal polarization states of a photon (Figure 1). Following the basics, students are guided through a series of questions in which there is no eavesdropper. These questions culminate in Table 1 that students complete which spans all possible combinations of Alice's polarization, Bob's polarization and whether or not Bob's detector clicks. In each situation, students are asked if a bit is recorded by Alice and Bob and if so which bit is generated (0 or 1). After completing this table in which students predict what should happen in each situation, they were provided with a figure that illustrates what should happen in a given situation (Figure 2). Figure 2 displays the probabilities that a bit is recorded in each case. Students are then guided to address cases in which Eve has eavesdropped on the key generation process. The questions culminate with students calculating that Eve will introduce an error in 25% of the bits that Bob records (in the ideal case with no decoherence, eavesdropping is the only source of error). They learn that if Alice and Bob compare a small subset of the shared key after the entire key is generated,

**Table 1.** In the QKD protocol in the QuILT, students are asked to complete the empty boxes in the table below predicting what should happen in a given situation based upon their understanding of the QKD protocol involving non-orthogonal polarization states of a photon.

| Alice's polarization: | ⤢ | ↔ | ⤢ | ↔ | ⤢ | ↔ |
|---|---|---|---|---|---|---|
| Bob's polarization: | ⤡ | ⤡ | ↕ | ↕ | ↕ | ⤡ |
| Bob's detector clicks: | N | N | Y | N | N | Y |
| Bit is recorded: (Y or N) | | | | | | |
| Which bit they record: (0 or 1 or -) | | | | | | |

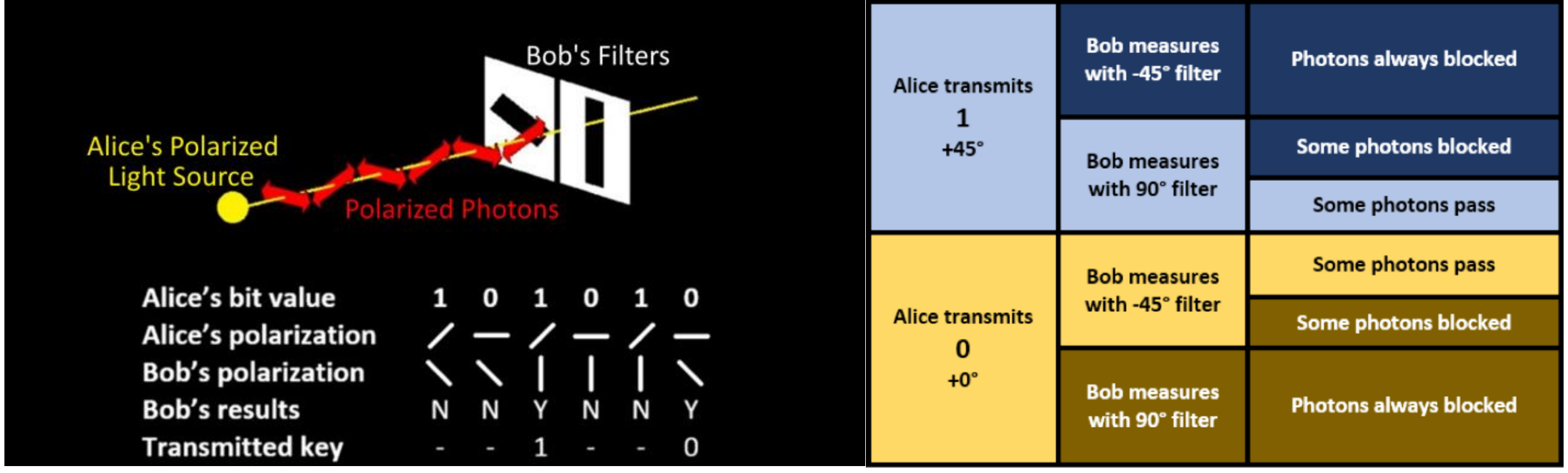

**Figure 2.** In the QKD QuILT, students are first asked to make predictions about what should happen in a given situation and later asked to check whether their predictions are consistent with the information in this figure about the QKD protocol involving non-orthogonal polarization states of single photons. Students are then asked to reconcile differences between their predictions and what happens in a given situation if their predictions are not consistent with this figure before they proceed.

they will know that this error is due to eavesdropping and they will discard the key and attempt the key generation process at another time, preferably through other channels. The following is an example of a question students answer as the culmination of a series of questions regarding Eve's interference in the key generation process:

Q: When Eve's detector does not click after she intercepts Alice's photon, she is not sure about the polarization of the photon sent by Alice. Eve will have to guess the polarization of the replacement photon she will send to Bob (in place of the one she intercepted). Suppose Eve is intercepting every photon sent by Alice and sending a replacement photon to Bob. Choose all of the following statements that are correct about the strategy for sending replacement photons to Bob when Eve is not sure what she intercepted from Alice:

(I) If Eve's polarizer is at 90° when she intercepted the photon sent by Alice and the detector does not click, she has less chance of making an error if she sends a replacement photon with 0° polarization to Bob.

(II) If Eve's polarizer is at -45° when she intercepted the photon sent by Alice and the detector does not click, she has less chance of making an error if she sends a replacement photon with 45° polarization to Bob.

(III) If Eve's polarizer is at 90° and the detector does not click, she has the same chance of making an error, regardless of whether she sends 0° or 45° polarization photon to Bob to replace it.

## 5. QUALITATIVE DETAILS OF STUDENT DIFFICULTIES

As noted, during the development of the QKD QuILT, we conducted 15 individual semi-structured think-aloud interviews [59] with physics undergraduate students enrolled in quantum mechanics and physics Ph.D. students who had taken graduate quantum mechanics to understand their difficulties with relevant concepts in order to effectively address them in the QuILT. Based upon these individual discussions during the interviews at different points of time in the QKD QuILT development, we made changes to the QuILT (earlier interviews led to larger changes compared to the later ones which were refinements). The qualitative details of the difficulties in this section are mainly based on these interviews and quantitative data based upon pretest and posttest will be presented in the next section even though we mention in this section that written pre/posttest data presented in the next section corroborate the interview findings. In particular, in this qualitative section, we do not provide the number of students who had each difficulty since the semi-structure interviews evolved with time as the different parts of the QKD QuILT were developed and fine-tuned and interviews generally progressed differently depending on how each student approached different questions. Also, all interviewed students were not probed about the same issues due to the time constraints. Later in the discussion of the written pretest and posttest, we focus on quantitative findings that relate to these issues in the context of student performance on each question.

During the interviews while developing the QuILT, some students claimed that the polarization states of a photon cannot be used as basis vectors for a two state system due to the fact that a photon can have infinitely many polarization states. They argued that since a polarizer can have any orientation and the orientation of the polarizer determines the polarization state of a photon incident on the polarizer, it did not make sense to think about polarization states of a photon as a two state system. These students were sometimes so fixated on their prior experiences with polarizers from introductory classes (which can be rotated to make their polarization axis whichever way one wants with respect to the polarization of incident light) that they had difficulty thinking of polarization states of a photon as vectors in a two-dimensional Hilbert space. It is interesting to note that students who had this difficulty often had no difficulty accepting that spin states of a spin-½ particle can be used as basis states for a two state system despite the fact that the two systems are analogous. Interviews suggest that this difference in their perception was often due to how a spin-½ system and polarization were first introduced to them and the kinds of mental models they had built about each of these systems. Generally, students are introduced to polarization in an introductory physics course in classical optics and they are introduced to spin-½ systems in a quantum mechanics course. Since students had learned about the spin-½ system only in quantum mechanics they had little difficulty thinking of them in terms of a two-dimensional Hilbert space.

As noted, some students who had difficulty connecting the measurement of polarization in a laboratory with polarization states of a photon were relatively comfortable with reasoning about measurement of a particular component of spin and thinking of spin states as vectors in a two-dimensional Hilbert space. During interviews, some even mentioned that measurement of $S_z$ will collapse a spin state which was initially in a superposition of eigenstates of $\hat{S}_z$ to an eigenstate of $\hat{S}_z$. However, they had difficulty with similar reasoning about the measurement of polarization of a photon collapsing the polarization state which is initially in a superposition state (e.g., the simplest case, in a superposition of two orthogonal polarization states chosen to be parallel and perpendicular to the polarization axis of the polarizer) into an eigenstate of the polarization measured. Even after students were reminded that a spin state of a spin-½ particle can be in a superposition of eigenstates of $\hat{S}_z$ and that eigenstates of $\hat{S}_z$ can be used as basis states to write any state in this two-dimensional Hilbert space, some interviewed students still had difficulty thinking of spin-½ and polarization states of photon as analogous and coming to terms with polarization states of a photon as vectors in a two-dimensional Hilbert space. They had difficulty with the notion that we can associate a vector in the Hilbert space to correspond to any polarization state (which is an eigenstate of the corresponding polarization) and that any two orthogonal states in that space can be used as basis states to represent any polarization state. Interviews suggest that the difference in how students were first introduced to spin and polarization was at least partly responsible for why they often reasoned about these analogous two-state systems in very different ways.

Some interviews and written responses (quantitative discussion is presented in the next section when discussing the pretest and posttest findings) suggest that some students incorrectly thought that whenever single photons with a given polarization were sent through a polarizer, they would be completely blocked (absorbed) by the polarizer only when the photon polarization was orthogonal to the polarization axis. These students thought that a photon with any other polarization would make the photodetector behind the polarizer click. Individual discussions suggest that this difficulty was often related to the difficulty in applying the measurement postulate of quantum mechanics to the single photon incident on a polarizer situation. In particular, some of these students thought that a single photon incident on a polarizer can partly pass through the polarizer and partly get absorbed (e.g., for polarized photons, they thought that the cosine squared of the angle between the polarization axis of the polarizer and polarization of the photon yields the fraction of a photon that transmits as opposed to the probability of the photon being transmitted). They did not realize that a single photon will either get absorbed by the polarizer or it will pass through with a certain probability. Interviews suggest that this difficulty often had its origin in the fact that students were not considering single photons passing through a polarizer probabilistically and were interpreting the situation by mixing quantum mechanical and classical ideas. In particular, students often confused the situation of a single photon incident on a polarizer with that of a beam of light incident on a polarizer. They knew from classical optics that the intensity of light generally decreases after passing through a polarizer and is only completely absorbed if the polarization of incident light is orthogonal to the polarization axis of the polarizer and extrapolated this, incorrectly, to the single photon case.

Examining holistically, the interviews and written responses also suggest that some students felt that the polarizer will only block photons that are polarized perpendicular to the axis of the polarizer and will allow any other photons to pass through completely. On the other hand, some students thought that the polarizer will act in the opposite extreme, blocking all photons that are not polarized parallel to the axis of the polarizer. In the written responses to be discussed later with quantitative results for the pretest and posttest, these two types of difficulties were less common but still serve to illustrate difficulties with the basics of how single photons interact with a polarizer, something that is important to help students understand the quantum mechanics principles at work in the QKD protocol.

One difficulty before working on the QKD protocol in the QuILT involves a lack of understanding of when both parties in the key generation process have successfully generated the next “bit” in the shared key in this protocol. Students with these difficulties did not realize that for each photon polarization that Alice sends, there is a non-zero probability of the photon being absorbed completely but only one of the two possible polarizations sent by Alice has a non-zero probability of being transmitted through each of Bob’s two polarizers that he randomly uses for his measurement. Often this difficulty resulted in students incorrectly assuming that it is possible for Bob to determine the polarization state of the photon Alice sent both if the photon is absorbed (not detected by the detector behind Bob’s polarizer) or transmitted (detected by the detector behind Bob’s polarizer). Students who had this difficulty often claimed that Bob will know the polarization of the photon if his detector does not click (as discussed earlier) due to their incorrect mental model that polarization of the photon must be perpendicular to the polarization axis of the polarizer for the photon to be completely absorbed. They also claimed that if the detector clicks they will know the polarization because a photon will be partly transmitted and partly absorbed (resulting in the detector clicking) in all situations in which the photon polarization is neither parallel nor orthogonal to the polarization axis of the polarizer. As a result of this difficulty, some students claimed that for all cases except when the photon polarization is perpendicular to Bob’s polarization axis in the given situation, Bob’s detector behind his polarizer would click. Due to these two difficulties, they incorrectly claimed that whether the detector clicks (signifying a photon passed through Bob’s polarizer) or not, the next “bit” of the key can be generated by the two parties.

Another way in which students sometimes misinterpreted when the next “bit” of the key is generated is that they thought that Bob is only certain about the polarization of the photon when the photon Alice sent is polarized perpendicular to the polarization axis of his polarizer. These students incorrectly claimed that since photons with polarization perpendicular to the polarization axis of Bob’s polarizer are always blocked, Bob must always know the polarization of the photon Alice sent when it is perpendicular to the polarization axis of the polarizer. In the case when photons are neither perpendicular nor parallel to the polarization axis of the polarizer, these students often claimed that since there is a non-zero chance for the photon to be either transmitted or absorbed, Bob cannot infer the polarization of the photon sent by Alice. Thus, these students incorrectly claimed that Bob can only generate the next bit of the key when there is only one possible outcome (when Bob's polarization axis is perpendicular to the polarization of the photon that Alice sends). They did not realize that what is important is whether a given outcome (transmitted or absorbed) exists for both polarizations of photon that Alice could send or only for one of the polarizations. Bob knows about the photon polarization only when his photodetector clicks because his photodetector may not click when photons are absorbed which can occur for either of the two polarizations of the photons that Alice sends.

Some students also had difficulty in distinguishing between a "qubit" and a "bit" and how a qubit can be in a superposition state but once a measurement of an observable is performed, we get one bit of information. Also, some had difficulty in determining which angle was relevant for determining if the photon will pass through or get absorbed by a polarizer. In particular, these students were unsure whether the relevant angles were those between the polarization of the incident photon and the polarization axis of the polarizer. For example, some students always made use of the polarization of the photon with respect to the horizontal axis in order to determine the probability of transmission or absorption incorrectly.

## 6. QUANTITATIVE RESULTS FROM THE PRETEST, POSTTEST, AND DELAYED POSTTEST

**Table 2.** Pretest, posttest, and delayed posttest results for the QKD QuILT broken down by question. The pretest and posttest results and corresponding *t*-tests and effect sizes using Cohen's d are based on 74 students' paired responses with an average score of 52.7% (pretest) and 90.8% (posttest), respectively. The delayed posttest results are based on answers from 24 students (with the associated Wilcoxon ranked-sum test and Cohen's d calculations performed with matched posttest and delayed posttest data). The two questions that make up the delayed posttest were selected as particularly good examples of questions that require an understanding of QKD and quantum mechanics principles related to the QKD protocol to answer. For each question involved in this table, the superscript "a" means $p < 0.05$, "b" means $p < 0.01$, and "c" means $p < 0.001$ for the *t*-test or Wilcoxon ranked-sum test conducted with the null hypothesis that there is no statistically significant difference between the means of pretest and posttest or posttest and delayed posttest, respectively. For pretest and posttest comparisons using *t*-test, *t*-values are reported with the critical *t*-value for $p < 0.05$ being 1.99 and for posttest and delayed posttest comparisons using Wilcoxon ranked-sum test, W is reported with the critical value of W for $p < 0.05$ being 492.

| Question # | % Correct on Pretest | % Correct on Posttest | t-value | Cohen's d | % Correct on Delayed Posttest | W | Cohen's d |
|---|---|---|---|---|---|---|---|
| 1 | 51.4 | $93.2^{c}$ | 7.25 | 1.06 | - | - | - |
| 2 | 63.5 | $91.9^{c}$ | 5.38 | 0.73 | 79.2 | 552 | 0.36 |
| 3 | 41.9 | $90.5^{c}$ | 8.32 | 1.20 | - | - | - |
| 4 | 63.5 | $98.6^{c}$ | 6.29 | 1.00 | - | - | - |
| 5 | 52.7 | $98.6^{c}$ | 7.88 | 1.27 | 83.3 | 540 | 0.63 |
| 6 | 48.6 | $78.4^{c}$ | 5.56 | 0.65 | - | - | - |
| 7 Box 1 | 71.6 | $97.3^{c}$ | 5.02 | 0.76 | - | - | - |
| 7 Box 2 | 45.9 | $95.9^{c}$ | 8.54 | 1.32 | - | - | - |
| 7 Box 3 | 47.3 | $93.2^{c}$ | 7.88 | 1.16 | - | - | - |
| 7 Box 4 | 79.7 | $98.6^{c}$ | 4.13 | 0.64 | - | - | - |
| 8 | 13.5 | $62.2^{c}$ | 8.32 | 1.16 | - | - | - |
| Average | 52.7 | $90.8^{c}$ | 10.2 | 0.92 | - | - | - |

**Results for each Pretest/Posttest Question**

Q1 in the Appendix evaluates students' understanding of when a bit of the key cannot be generated. Since the detector does not click, Bob cannot infer anything with certainty about the polarization of the photon Alice sent and thus cannot use this photon to record the next bit making the correct answer "None of the above". Examining the pretest data, we note that roughly 50% of students incorrectly claimed that a "bit" of the key can be generated even when no photon is detected. After working on the QuILT, practically no students had this difficulty. The major difficulty that was common in this question is that some students incorrectly claimed that the photon that Alice sent is polarized perpendicular to Bob's polarizer and Bob is 100% sure about the polarization of the photon sent by Alice.

Q2 in the Appendix supplements the first question. It evaluates student understanding of when a "bit" of the key can be generated. When the photodetector clicks, Bob knows that the photon that Alice sent cannot be polarized perpendicular to Bob's polarizer and therefore he can determine the polarization of the photon he detected, so that the correct answer is "(I) and (III) only". The number of students who answered this question correctly on the pretest is likely to be somewhat inflated due to one of the common incorrect models that students used to describe QKD that leads students to select the correct answer due to incorrect reasoning. These students thought that the polarizer only blocked photons that were polarized perpendicular to its axis and that all photons that were not perpendicular to its axis were allowed to pass through (either fully or partially) causing the detector to click. Thus, they would answer that Bob knows the polarization of the photon that Alice sent and that the photon has the polarization which is not perpendicular to the axis of the polarizer that Bob used. Though these students made the correct selection from the options given, they incorrectly thought that this setup will always result in Bob's detector clicking. Table 2 shows that about 40% of students failed to answer this question correctly on the pretest but very few student had this difficulty after working on the QuILT. Even in the delayed posttest after two months, student performance is substantially better than the pretest.

Q3 in the Appendix evaluates student understanding of how a polarizer interacts with single polarized photons. Students must know that the probability of a photon passing through a polarizer is $\cos^2\theta$ where $\theta$ is the difference between the polarization of the photon and the polarization axis of the polarizer ($\cos^2 20°$). Two common student difficulties became apparent on the pretest. The first involved students using $\sin^2\theta$ rather than $\cos^2\theta$ which suggests a lack of understanding of the transmission probability through a polarizer for a polarized photon. The second difficulty was identifying the angle in the equation for transmission probability incorrectly. Many students selected one of the angles provided in the problem statement rather than taking the difference between photon polarization and the polarization axis of Bob's polarizer. These two difficulties and other less common mistakes resulted in more than half of the students failing to correctly answer this question on the pretest. After the QuILT, very few students made a mistake on this question.

Q4 in the Appendix investigates student understanding of the likelihood that Bob's polarizer would block the photon with a given polarization. Question 4 closely parallels question 3 and requires that students calculate the probability of a photon passing through the polarizer to be roughly 88.3%. Similarly to question 3, there are two major difficulties that students have with this question. One difficulty which was discussed earlier stems from using $\sin^2\theta$ rather than $\cos^2\theta$ to predict the probability of transmission. The second difficulty involves assuming that the photon is either always transmitted or always absorbed despite the fact that photon polarization and the polarization axis of Bob's polarizer are neither orthogonal nor parallel. These difficulties resulted in roughly 40% of students answering this question incorrectly in pretest. Table 2 shows that after working on the QuILT, practically none of the students answered this question incorrectly.

Q5 in the Appendix examines concepts similar to those examined in Question 1. The primary concept emphasized in this question involves understanding that the photon's polarization state, and thus the next "bit" of information, can be determined only when the detector clicks, so that the correct answer is option c. Two of the available options assert that the polarization of the photon that Alice sent can be determined and is one of the two possible polarization angles. A third option, "none of the above", is logically unsound because the question statement gives every piece of information that Bob has in this situation meaning that he must be able to make some sort of determination, as he would during key generation. These three distracting options lead to roughly half of students selecting the incorrect answers in the pretest. The most common choice is based on the same common difficulty that students had with question 1. These students claimed that the photon is only blocked when the polarization of the photon is perpendicular to the axis of the polarizer and therefore they select the option that states that Bob can identify the polarization of the photon and that it is polarized perpendicular to the axis of his polarizer. After working on the QuILT, almost all students answer this question correctly. Table 2 shows that after two months slightly over 15% of students answered this question incorrectly.

Q6 in the Appendix evaluates student understanding of several important concepts relevant for the QKD protocol. It simultaneously tests whether students understand that if Bob's detector clicks, the polarization state of Alice's photon can be determined and thus the next bit of the key can be generated and if the detector does not click, the polarization state of the photon Alice sent cannot be inferred. Thus, the correct answer is "(I) and (II) only". This question also includes a good distractor related to a common difficulty students have before the QuILT, namely, that a single photon can be partly absorbed and partly transmitted (as discussed in the difficulty section earlier, this difficulty was often a result of over generalization of the fact that when a beam of light passes through a polarizer, the intensity of light generally decreases due to some light getting absorbed and some passing through). About half of the students selected the correct answer in the pretest. The two most common incorrect answers are based on two similar difficulties discussed earlier. Both difficulties involve Bob being able to determine the polarization of the photon that Alice sent regardless of whether the detector clicks or not. Students were often confused about what happens to photons that are polarized neither perpendicular nor parallel to the axis of the polarizer. They incorrectly thought that the photon is either partially transmitted and partially absorbed resulting in the transmitted portion of the photon making the detector click or that they are completely transmitted allowing the detector to click. Table 2 shows that after the QuILT, roughly 80% of students correctly answered this question.

Since Q7 in the Appendix only requires students to take the difference between the two angles and use it with the equation for the probability that the photon will be transmitted ($\cos^2\theta$, which is 0%, approximately 88.3%, 88.3% and 0%, respectively) it is reasonable to expect some uniformity across these four questions. Upon examining the pretest data in Table 2, we observe that the percentage of students who correctly answered each sub-question ranges from roughly 50% to 80%. One reason for the range of correct answers is that many students did not know the expression for the probability that the photon will be transmitted, but they knew that a photon will not pass through an orthogonal polarizer. After working on the QuILT, nearly all of the students answered these four questions correctly.

The primary difficulty on Q8 in the Appendix was to properly account for each of the four probabilities in the table to determine the overall probability of roughly 44.2%. In particular, several students ignored any probabilities from the table that were equal to zero and proceeded to average the remaining probabilities. Other students simply wrote down 50% due to there being a nonzero probability of generating a bit of the key 50% of the time. A slightly less common error that several students made involved multiplying the nonzero probabilities together to determine the average probability of generating a shared bit of the key. Table 2 shows that slightly over 10% of students answered this question correctly on the pretest while roughly 60% of students determined this probability correctly on the posttest.

**Overall Results for the Pretest, Posttest, and Delayed Posttest**

The pretest and posttest responses suggest that students often have alternative models of concepts relevant for QKD and they have difficulty with various concepts including the polarization states of a single photon and the single photon state upon measurement. Largely these models are consistent with the issues discussed in the student difficulties section. These alternative models are predominantly observed in student responses on the pretest and are mostly reconciled by the time students take the posttest after engaging with the QKD QuILT.

Examination of students' responses (especially across the pretest questions) reveals inconsistencies in student reasoning. One set of inconsistencies is displayed by student responses to Questions 3, 4 and 7. Despite the fact that Questions 3 and 4 are nearly identical questions and are both contained within the answers to Question 7, it was common on the pretest to find an answer for one of these three questions that did not match the pattern being followed in the other two questions and in some cases the boxes for Questions 7 that were supposed to be similar to responses to Questions 3 and 4 had different student responses. Another common inconsistency is found when examining the responses to Questions 1 and 5 on the pretest. Despite these two questions being nearly identical to each other in subject matter, students often selected inconsistent responses to these questions. On the pretest, even on other questions student reasoning was often inconsistent.

Interviews suggest that one cause of inconsistency across answers is the context of the questions. For example, when students are asked questions that require conceptual answers and require no equations or mathematical manipulations (e.g., Questions 1, 2, 5 etc.), they are more likely to answer them based upon their alternative model. If instead they are asked a question that requires a mathematical answer, e.g., the probability of a photon passing through a polarizer (e.g., Questions 3, 4, 7 etc.), students are likely to rely on equations even if the results disagree with their model (students may not even compare these answers with their model to check for consistency). Other subtle changes in the context also affect student responses to similar questions. When comparing Questions 1 and 5, the major difference is the change in the layout of the available answers but it was enough to cause students to answer these two questions inconsistently, especially on the pretest. Interviews suggest that some students on the pretest became overly focused on the details of each individual question. In the more extreme cases, this was taken to the point where students answered each question without thinking about its interrelation with other questions. This type of treatment was at least partly responsible for students answering Questions 3 and 4 differently than Question 7 (which requires the answers to Questions 3 and 4). Students became too focused on each individual question and did not check for consistency across questions.

Students' posttest responses are markedly more consistent across questions. T-tests were performed comparing each pretest question with the corresponding posttest problem showing statistically significant ($p < 0.001$) improvements on all questions, as shown in Table 2. Interviews and written responses suggest that the QKD QuILT also improves student understanding of polarization states of a photon and how to make connection between a polarizer in a physical space and polarization states of a photon in a Hilbert space. It also helps students think about single photons as having a probability of transmission or absorption rather than being partially transmitted and partially absorbed. The QuILT helps students learn that quantum measurement collapses the polarization state of a photon depending upon the polarizer used. Overall, the QKD QuILT was helpful in improving student understanding of the relevant topics.

The delayed posttest results show that despite the scores not remaining as high as they are for the posttest given immediately after the QuILT they are markedly higher than student scores on the pretest for both of the questions examined. Wilcoxon ranked-sum tests were performed comparing the corresponding posttest questions to the two delayed posttest problems and the results show no statistically significant differences existed between the posttest and the delayed posttest for either question, as shown in Table 2. The two questions that make up the delayed posttest were selected as particularly good examples of questions that require an understanding of QKD and quantum mechanics principles related to the the QKD protocol to answer.

## 7. SUMMARY

We developed, validated and evaluated a QuILT to help upper-level undergraduate students learn quantum cryptography in the context of quantum key distribution using a protocol that involves non-orthogonal polarization states of photons [22]. The secure QKD protocol that students learn in the QuILT focuses on generation of a shared key securely over a public channel using two non-orthogonal polarization states of single photons. The QKD QuILT provides a guided approach to bridge the gap between the quantitative and conceptual aspects of quantum mechanics. It keeps students actively engaged in the learning process and in-class evaluation shows that the QuILT improves students' understanding of concepts related to the QKD. Performance on the QuILT pretest showed that students were able to answer only slightly over half of the questions about the QKD protocol after a typical lecture-based treatment of the quantum mechanics principles related to this protocol. After working through the QuILT in the class, student understanding of the QKD protocol and the associated physics principles were improved such that the average score was roughly 90% on the posttest which is encouraging.

Although some undergraduate lab experiments have been developed to help students learn quantum mechanics using single photon experiments, e.g., using Mach Zehnder interferometer [7], to our knowledge no research-validated learning tools have been disseminated via publication for helping students learn QKD. QKD involves an exciting real world application of quantum mechanics in the context of quantum cryptography that can be taught to students in an upper-level undergraduate quantum mechanics course. Considering the potential of these types of learning tools in helping students learn about this rapidly evolving field of quantum information science and technology, more research-validated learning tools are needed in this area to help students with diverse backgrounds master these challenging concepts. We are currently in the process of implementing another QKD QuILT in upper-level quantum mechanics course that uses two entangled spin-1/2 particles that has been validated outside of class iteratively using a process similar to that described here for the QKD QuILT involving two non-orthogonal polarization states of light. Individual interviews with students suggest that entanglement is a very challenging concept for students so the QKD QuILT involving two entangled spin-1/2 particles is comparable in difficulties to the QKD QuILT involving non-orthogonal polarization states of photons described here. Finally, we note that we continue to develop and validate other learning and assessment tools to help instructors teach quantum mechanics effectively [62-65].

## ACKNOWLEDGEMENTS

We thank Qin Xu and Chris Schunn for helpful discussions. We thank the US National Science Foundation for award PHY-1806691.

# REFERENCES

1. P. Jolly, D. Zollman, S. Rebello, and A. Dimitrova, Visualizing potential energy diagrams, Am. J. Phys. **66**(1), 57 (1998).
2. C. Singh, Student understanding of quantum mechanics, Am. J. Phys. **69** (8), 885 (2001).
3. D. Zollman, N. S. Rebello, and K. Hogg, Quantum mechanics for everyone: Hands-on activities integrated with technology, Am. J. Phys. **70**, 252 (2002).
4. R. Muller and H. Wiesner, Teaching quantum mechanics on an introductory level, Am. J. Phys. **70**, 200 (2002).
5. M. Wittmann, R. Steinberg et al., Investigating student understanding of quantum physics: Spontaneous models of conductivity, Am. J. Phys. **70**, 218 (2002).
6. D. Domert, C. Linder and A. Ingerman, Probability as a conceptual hurdle to understanding one-dimensional quantum scattering and tunneling, Eur. J. Phys. 26, 47 (2005).
7. E. Galvez and C. Holbrow, Interference with correlated photons: Five quantum mechanics experiments for undergraduates, Am. J. Phys. **73**, 127 (2005).
8. C. Singh, Student understanding of quantum mechanics at the beginning of graduate instruction, Am. J. Phys. **76**, 277 (2008).
9. C. Manogue, E. Gire, D. McIntyre and J. Tate, Representations for a Spins-First approach to quantum mechanics, Proc. Physics Education Research Conf., 2011 (doi: 10.1063/1.3679992)
10. A. Kohnle et al., A new introductory quantum mechanics curriculum, Eur. J. Phys. **35**, 015001 (2014).
11. E. Gire and E. Price, The structural features of algebraic quantum notations, Phys. Rev. ST PER **11**, 020109 (2015).
12. G. Passante et al., Examining student ideas about energy measurements on quantum states across undergraduate and graduate level, Phys. Rev. ST Phys. Educ. Res. **11**, 020111 (2015).
13. P. Emigh et al., Student understanding of time dependence in quantum mechanics, Phys. Rev. ST Phys. Educ. Res. **11**, 020112 (2015).
14. V. Dini and D. Hammer, Case study of a successful learner's epistemological framings of quantum mechanics, Phys. Rev. Phys. Educ. Res. **13**, 010124 (2017).
15. C. Porter and A. Heckler, Graduate student misunderstandings of wave functions in an asymmetric well, Phys. Rev. PER **15**, 010139 (2019).
16. C. Singh, Transfer of learning in quantum mechanics, in Proc. Phys. Educ. Res. Conf. **790**, p. 23 (2005) (http://aip.scitation.org/doi/abs/10.1063/1.2084692)
17. C. Singh, Assessing and improving student understanding of quantum mechanics, in Proc. Phys. Educ. Res. Conf. **818**, p. 69 (2006) (http://aip.scitation.org/doi/abs/10.1063/1.2177025)
18. C. Singh, M. Belloni, and W. Christian, Improving students' understanding of quantum mechanics, Phys. Today **59** (8), 43-49 (2006).
19. C. Singh, Student difficulties with quantum mechanics formalism, in Proc. Phys. Educ. Res. Conf. **883**, p. 185 (2007) (http://aip.scitation.org/doi/abs/10.1063/1.2508723)
20. C. Singh, Helping students learn quantum mechanics for quantum computing, in Proc. Phys. Educ. Res. Conf. **883**, p. 42 AIP Conf. Proc. (2007) (http://aip.scitation.org/doi/abs/10.1063/1.2508687)
21. C. Singh, Interactive learning tutorials on quantum mechanics, Am. J. Phys. **76** (4), 400-405 (2008).
22. C. Singh and G. Zhu, Cognitive issues in learning advanced physics: An example from quantum mechanics, in Proc. Phys. Educ. Res. Conf. 1179, p. 63 (2009), (http://doi.org/10.1063/1.3266755).
23. A. J. Mason and C. Singh, Do advanced students learn from their mistakes without explicit intervention?, Am. J. Phys. **78**, 760-767 (2010).
24. S. Y. Lin and C. Singh, Categorization of quantum mechanics problems by professors and students, Euro. J. Phys. **31**, 57-68 (2010).
25. G. Zhu and C. Singh, Improving students' understanding of quantum mechanics via the Stern-Gerlach experiment, Am. J. Phys. **79** (5), 499-507 (2011).

26. C. Singh and G. Zhu, Improving students' understanding of quantum mechanics by using peer instruction tools, in Proc. Phys. Educ. Res. Conf. **1413**, p. 77 (2012), (http://doi.org/10.1063/1.3679998).
27. G. Zhu and C. Singh, Surveying students' understanding of quantum mechanics in one spatial dimension, Am. J. Phys. **80** (3), 252-259 (2012).
28. G. Zhu and C. Singh, Improving students' understanding of quantum measurement: I. Investigation of difficulties, Phys. Rev. ST PER **8** (1), 010117 (2012).
29. G. Zhu and C. Singh, Improving students' understanding of quantum measurement: II. Development of research-based learning tools, Phys. Rev. ST PER **8** (1), 010118 (2012).
30. G. Zhu and C. Singh, Improving student understanding of addition of angular momentum in quantum mechanics, Phys. Rev. ST PER **9**, 010101 (2013).
31. C. Singh and E. Marshman, Investigating student difficulties with Dirac notation, in Proc. Phys. Educ. Res. Conf. p. 345 (2014) (https://www.compadre.org/per/items/detail.cfm?ID=13149)
32. S. DeVore and C. Singh, Development of an interactive tutorial on quantum key distribution, in Proc. Phys. Educ. Res. Conf. p. 59 (2015), (https://www.compadre.org/per/items/detail.cfm?ID=13448).
33. B. Brown and C. Singh, Development and evaluation of a quantum interactive learning tutorial on Larmor Precession of spin, in Proc. Phys. Educ. Res. Conf. p. 47 (2015), (http://doi.org/10.1119/perc.2014.pr.008).
34. C. Singh and E. Marshman, Review of student difficulties in upper-level quantum mechanics, Phys. Rev. ST PER **11**, 020117 (2015).
35. E. Marshman and C. Singh, Framework for understanding the patterns of student difficulties in quantum mechanics, Phys. Rev. ST PER **11**, 020119 (2015).
36. B. Brown, A. Mason, and C. Singh, Improving performance in quantum mechanics with explicit incentives to correct mistakes, Phys. Rev. PER 12, 010121 (2016).
37. E. Marshman and C. Singh, Investigating and improving student understanding of quantum mechanical observables and their corresponding operators in Dirac notation, Eur. J. Phys. **39**, 015707 (2017).
38. E. Marshman and C. Singh, Investigating and improving student understanding of the probability distributions for measuring physical observables in quantum mechanics, Euro. J. Phys. **38**, 025705 (2017).
39. E. Marshman and C. Singh, Investigating and improving student understanding of the expectation values of observables in quantum mechanics, Eur. J. Phys. **38**, 045701 (2017).
40. R. Sayer, A. Maries and C. Singh, A quantum interactive learning tutorial on the double-slit experiment to improve student understanding of quantum mechanics, Phys Rev PER 13, 010123 (2017).
41. S. Siddiqui and C. Singh, How diverse are physics instructors' attitudes and approaches to teaching undergraduate-level quantum mechanics?, Eur. J. Phys. **38**, 035703 (2017).
42. E. Marshman and C. Singh, Investigating and improving student understanding of quantum mechanics in the context of single photon interference, Phys. Rev. PER **13**, 010117 (2017).
43. E. Marshman and C. Singh, Interactive tutorial to improve student understanding of single photon experiments involving a Mach-Zehnder interferometer, Eur. J. Phys. **37**, 024001 (2016).
44. A. Maries, R. Sayer and C. Singh, Effectiveness of interactive tutorials in promoting "which-path" information reasoning in advanced quantum mechanics, Phys. Rev. PER **13**, 020115 (2017).
45. C. Keebaugh, E. Marshman and C. Singh, Investigating and addressing student difficulties with a good basis for finding perturbative corrections in the context of degenerate perturbation theory, Eur. J. Phys. **39,** 055701 (2018).
46. C. Keebaugh, E. Marshman and C. Singh, Improving student understanding of fine structure corrections to the energy spectrum of the hydrogen atom, Am. J. Phys. **87**, 594 (2019).
47. C. Keebaugh, E. Marshman and C. Singh, Investigating and addressing student difficulties with the corrections to the energies of the hydrogen atom for the strong and weak field Zeeman effect, Eur. J. Phys. **39**, 045701 (2018).
48. C. Keebaugh, E. Marshman and C. Singh, Improving student understanding of a system of identical particles with a fixed total energy, Am. J. Phys. **87**, 583 (2019).
49. C. H. Bennett and G. Brassard, Quantum Cryptography: Public key distribution and coin tossing in *Proceedings of the IEEE International Conference on Computers, Systems, and Signal Processing*, Bangalore, 175 NY (1984).
50. C. Bennett, Quantum cryptography using any two nonorthogonal states, Phys. Rev. Lett. **68**, 3121-3124 (1992).

51. C. H. Bennett, G. Brassard, and N. D. Mermin, Quantum cryptography without Bell's theorem, Phys. Rev. Lett. **68**(5), 557 (1992).
52. I. Gerhardt, et al. Full-field implementation of a perfect eavesdropper on a quantum cryptography system, Nature *Communications* **2**, 1 (2011).
53. C. H. Bennett, F. Bessette, G. Brassard, L. Salvail, and J. Smolin, Experimental quantum cryptography, J. of Cryptology, **5**(1), 3 (1992); G. Brassard and L. Salvail, Secret key reconciliation by public discussion, Advances in Cryptology: Euro crypt 93 *Proc*. 410 (1993).
54. R. Hughes and J. Nordholt, Refining quantum cryptography, Science **333**(6049), 1584 (2011).
55. A. Kohnle, Quantum Cryptography, QuVis: The University of St. Andrews Quantum Mechanics Visualisation project. University of St. Andrews, http://www.st-andrews.ac.uk/physics/quvis. See http://www.st-andrews.ac.uk/physics/quvis/simulations_html5/sims/cryptography-b92/B92_photons.html
56. An example of a company that performs quantum cryptography for the financial industry can be found at http://www.idquantique.com/
57. R. Alleaume, SECOQC White Paper on quantum key distribution and cryptography, SECOQC 3 (2007).
58. W. K. Wootters and W. H. Zurek, A single quantum cannot be cloned, Nature **299**, 802 (1982).
59. M. T. H. Chi, in The Think Aloud Method: A Practical Guide to Modeling Cognitive Processes, edited by Maarten W. van Someren, Yvonne F. Barnard, and Jacobijn A. C. Sandberg (Academic Press, London, 1994), pp. 1-12.
60. The entire QKD QuILT is available on the QuILTbeta website on COMPADRE http://www.opensourcephysics.org/items/detail.cfm?ID=13514
61. B. Kirwan and L. Ainsworth (Eds.), A Guide to Task Analysis, Taylor and Francis (1992).
62. C. Keebaugh, E. Marshman and C. Singh, Improving student understanding of corrections to the energy spectrum of the hydrogen atom for the Zeeman effect, Phys. Rev. Phys. Educ. Res. 15, 010113 (2018).
63. E. Marshman and C. Singh, Validation and administration of a conceptual survey on the formalism and postulates of quantum mechanics, Phys. Rev. PER 15, 020128 (2019).
64. P. Justice, E. Marshman and C. Singh, Improving student understanding of quantum mechanics underlying the Stern-Gerlach experiment using a research-validated multiple-choice question sequence, Eur. J. Phys. 40, 055702 (2019).
65. P. Justice, E. Marshman and C. Singh, Student understanding of Fermi energy, the Fermi-Dirac distribution and total electronic energy of a free electron gas, Eur. J. Phys. 41,015704 (2020).

## APPENDIX

The following are the questions in the pretest (posttest questions were similar with changes to the angles as discussed in the main section).

**Q1**: Bob uses the polarizer with a $-20°$ polarization angle and his photodetector does not click. Choose all of the following statements that can be inferred based upon the above protocol used by them:

a. Bob is 100% sure about the polarization of the photon sent by Alice.
b. Alice must have sent a photon with a $70°$ polarization.
c. Alice must have sent a photon with a $0°$ polarization.
d. None of the above.

**Q2**: Bob uses the polarizer with a $-20°$ polarization angle and his photodetector clicks. Choose all of the following statements that can be inferred based upon the above protocol used by Alice and Bob:

a. Bob is 100% sure about the polarization of the photon sent by Alice.
b. Alice must have sent photon with a $70°$ polarization.
c. Alice must have sent photon with a $0°$ polarization.

**Q3**: Suppose Alice transmits a photon with $70°$ polarization. Bob uses a polarizer with $90°$ polarization angle to intercept it. Write down the probability that the photon will pass through Bob's polarizer.

**Q4**: Alice transmits a photon with $0^\circ$ polarization and Bob uses a polarizer with $-20^\circ$ polarization angle. Which one of the following statements is true?

a. The photon is blocked by Bob's polarizer with a 100% certainty.
b. The photon will pass through Bob's polarizer with approximately 88.3% likelihood.
c. The photon will pass through Bob's polarizer with approximately 11.7% likelihood.
d. The photon will pass through Bob's polarizer with a 100% certainty.

**Q5**: Bob uses a polarizer with $90^\circ$ polarization angle and the detector does not click. Can he infer the polarization state of the photon that Alice sent? If so, what is it?

a. Yes. Alice must have sent a photon with $0^\circ$ polarization.
b. Yes. Alice must have sent a photon with $70^\circ$ polarization.
c. No. Alice could have sent a photon with either polarization ($0^\circ$ or $70^\circ$).
d. None of the above.

**Q6**: Choose all of the following statements that are correct based upon the protocol described:

a. Whenever Bob's detector clicks, he can infer the polarization of the photon that Alice sent.
b. Whenever Bob's detector does not click, he cannot infer the polarization of the photon that Alice sent.
c. If Alice sends a photon with $0^\circ$ polarization and Bob uses a polarizer with $-20^\circ$ polarization angle, that photon will be partly absorbed and partly transmitted.

**Q7**: Complete the third column of the following table by recording "the probability that a photon will pass through and hence Bob's detector clicks":

| | | Probability of detector clicking |
|---|---|---|
| Alice transmits $70^\circ$ | Bob uses $-20^\circ$ polarizer | |
| | Bob uses $90^\circ$ polarizer | |
| Alice transmits $0^\circ$ | Bob uses $-20^\circ$ polarizer | |
| | Bob uses $90^\circ$ polarizer | |

**Q8**: Using the table above for the case described in the preceding question, calculate the percentage of measurements in which Bob is 100% sure about the polarization of the photon that Alice sent out of all of the experiments that Alice and Bob conduct.